\documentstyle[12pt,aaspp4,flushrt]{article}
\def\lsim{\mathrel{\mathpalette\Oversim<}}
\def\gsim{\mathrel{\mathpalette\Oversim>}}
\def\Oversim#1#2{\lower0.5ex\vbox{\baselineskip0pt\lineskip0pt%
            \lineskiplimit0pt\ialign{%
          $\mathsurround0pt #1\hfil##\hfil$\crcr#2\crcr\sim\crcr}}}

\begin{document}
\title{
Formation of Primordial Galaxies under UV background Radiation
}
\author{Hajime Susa\altaffilmark{1} and 
Masayuki Umemura\altaffilmark{2}}
\affil{Center for Computational Physics, University of
  Tsukuba, Tsukuba 305, Japan }
\altaffiltext{1}{e-mail:susa@rccp.tsukuba.ac.jp}
\altaffiltext{2}{e-mail:umemura@rccp.tsukuba.ac.jp}
\begin{abstract}
The pancake collapse of pregalactic clouds
under UV background radiation is explored with a one-dimensional 
sheet model. 
Here, attention is concentrated on elucidating the basic physics
on the thermal evolution of pregalactic clouds exposed to diffuse UV
radiation.
So, we treat accurately the radiation transfer for the ionizing 
photons, with solving chemical reactions regarding
hydrogen molecules as well as atoms.
The self-shielding against UV radiation by H$_2$ Lyman-Werner bands,
which regulates the photo-dissociation of hydrogen molecules, 
is also taken into account.
As a result, it is found that when the 
UV background radiation is at a level of 
$10^{-22} (\nu/\nu_L)^{-1} 
{\rm erg~ s^{-1} cm^{-2} Hz^{-1} str^{-1}}$, 
the cloud evolution bifurcates with a critical mass as
$M_{\rm SB} = 2.2\times 10^{11} M_\odot \left[(1+z_c)/5\right]^{-4.2}$,
where $z_c$ is the final collapse epoch.
A cloud more massive than $M_{\rm SB}$ 
cools below $5\times 10^3$K due to H$_2$ line
emission at the pancake collapse and would undergo the initial starburst.
The pancake 
possibly evolves into a virialized system in a dissipationless fashion.
Consequently, this leads to the dissipationless galaxy formation 
at $3\lsim z_c \lsim 10$.
A cloud less massive than $M_{\rm SB}$ cannot cool by H$_2$ emission 
shortly after the pancake collapse, but could cool 
in the course of shrinking to the rotation barrier.
This is likely to lead to the dissipational galaxy formation
at relatively low redshifts as $0\lsim z_c \lsim 4$.
The present results provide a solid physical mechanism which controls
the star formation efficiency in the pregalactic clouds.
In the context of a standard CDM cosmology, 
$M_{\rm SB}$ lies between 1$\sigma$ and 2$\sigma$ 
density fluctuations. 
\end{abstract}
\newpage
\section{INTRODUCTION}
\label{intro}

It has been widely accepted that the formation of the first generation 
of objects, say Pop III objects, 
is regulated by the cooling by primordial hydrogen molecules.
A number of authors have explored the formation of Pop III
objects in so-called 'dark age' ($z>5$) by concentrating on
the role of hydrogen molecules 
(Matsuda, Sato, \& Takeda \markcite{MST65} 1965; Yoneyama \markcite{YO72}1972;
Hutchins \markcite{HU76}1976; Carlberg
\markcite{Car81} 1981; Palla, Salpeter, \& Stahler \markcite{PSS83}1983;
Susa, Uehara, \& Nishi \markcite{SUN96} 1996; 
Uehara et al.\markcite{USNY96} 1996; Annonis \& Norman \markcite{AN96} 1996; 
Tegmark et al. \markcite{TSR97} 1997; Nakamura \& Umemura \markcite{NU99}1999).
Also, even if the gas contains metals,
the metallic line cooling is overwhelmed by
the hydrogen/helium cooling at $T>10^4$K  
when the metallicity is lower than $10^{-2} Z_\odot$
(B\"{o}hringer \& Hensler \markcite{BH89}1989).
In fact, recently the metal abundance in the intergalactic space
is inferred to be at a level of $10^{-3} Z_\odot$ 
(Cowie et al. \markcite{Cow95}1995; Songaila \& Cowie \markcite{SC96}1996; 
Songaila \markcite{Son97}1997; Cowie \& Songaila\markcite{CS98} 1998). 
Moreover, at lower temperature of $T<10^4$K, 
the ${\rm H_2}$ cooling is estimated to be still predominant as long as 
the metallicity is lower than $10^{-2} Z_\odot$,
from the comparison of the cooling of the solar abundance gas 
(Spitzer 1978) with the maximal ${\rm H_2}$ cooling 
(Kang \& Shapiro 1992).
Hence, hydrogen molecules are likely to play a fundamental role on
the formation of galaxies under the metal poor environment.

In addition, there seem to be situations where
UV background radiation significantly influences the dynamical as well as 
thermal evolution of pregalactic clouds.
At the epochs of $z<5$, 
the existence of UV background of 
$10^{-21\pm 0.5}{\rm erg\; s^{-1} cm^{-2} str^{-1} Hz^{-1}}$
at the Lyman limit is inferred from so-called proximity effect of 
Ly$\alpha$ forest 
(Bajtlik, Duncan, \& Ostriker \markcite{BDO88}1988; 
Giallongo et al.\markcite{GCD96} 1996).
Also, the Gunn-Peterson optical depths
show that the intergalactic space was in fact highly ionized
(Gunn \& Peterson \markcite{GP65}1965; Schneider, Schmidt, \& Gunn \markcite{SSG89}1989, \markcite{SSG91}1991). 
The UV background is attributed to the radiation from
quasars and partially from young galaxies (e.g. Giallongo et al.\markcite{GDO96} 1996). 
Even in the dark age of the universe, say $z>5$, 
the first generation of objects as well as unseen proto-quasars
might be external photoionization sources for subsequently collapsing 
objects.
So far, many authors have considered the dynamical effects 
produced by the UV background radiation
(Umemura \& Ikeuchi\markcite{UI84} 1984; Dekel \& Rees\markcite{DR87} 1987; Babul \& Rees \markcite{BR92}1992; 
Efstathiou \markcite{Efs92}1992; Chiba \& Nath \markcite{CN94}1994; Thoul \& Weinberg \markcite{TW96}1996; Quinn, 
Katz, \& Efstathiou \markcite{QKE96}1996; 
Navarro \& Steinmetz \markcite{NS97}1997),
to account for Ly$\alpha$ clouds or to reconcile the paradox
that in the hierarchical clustering paradigm for galaxy 
formation, low mass galaxies are overproduced compared with 
observations (White \& Frenk \markcite{WF91}1991; Kauffman, White, \& Guiderdoni\markcite{KWG93} 1993;
Cole et al. \markcite{Cole94}1994). 
The simulations have been hitherto based upon the assumption that 
the medium is optically thin against UV photons. 
However, the gas clouds become optically thick as the gravitational
collapse proceeds. The significance of the optical depth has been stressed
by the accurate treatment on radiative transfer 
(Tajiri \& Umemura \markcite{TU98}1998).

The primary effects of UV radiation are
photoionization, UV heating, and photo-dissociation of
${\rm H_2}$ [Haiman, Rees, \& Loeb \markcite{HRL96}1996 (hereafter HRL), 
1997; Corbelli, Galli, \& Palla \markcite{CGP97}1997; Kepner, Babul, \& Spergel\markcite{KBS97} 1997]. 
HRL, with including radiation transfer effects in one-zone clouds,
estimated the UV heating rate and 
the cooling rate under the assumption of the chemical equilibrium.
They found that the cooling rate can exceed the heating rate at
$T\lsim 10^4 {\rm K}$ for high density. However, basically
the chemical reactions regarding ${\rm H_2}$ formation are not in
equilibrium (e.g. Susa et al. \markcite{SUNY98}1998; Bertoldi \& Draine
\markcite{BD96}1996; Diaz-Miller, Franco \& Shore \markcite{DFS}1998;
Hollenbach \& Tielens \markcite{HT99}1999).
In particular, if ionization processes such as  
shock heating or photoionization take place, the chemical processes
become out of equilibrium (e.g. Shapiro \& Kang \markcite{SK87}1987).
Thus, the non-equilibrium process for the formation of ${\rm H_2}$
should be involved, coupled with hydrodynamical process.

In this paper, we re-examine the thermal and dynamical evolution
of pregalactic clouds under UV background radiation.
In the present analysis, we solve properly 
the radiative transfer of diffuse 
UV photons and include non-equilibrium
chemical reactions regarding ${\rm H_2}$ formation due to
shock-ionization and photoionization. 
The goal is to elucidate the key physics on the effects  
of the UV background radiation upon the thermal evolution 
of pregalactic clouds. For the purpose,
we make the dynamical model as simple as possible.
A cosmological density perturbation far beyond the Jeans scale 
forms a flat pancake-like disk.
Although the pancake formation is originally studied 
by Zel'dovich \markcite{Zel70}(1970)
in the context of the adiabatic fluctuations in baryon or
hot dark matter-dominated universes, 
recent numerical simulations have shown that 
such pancake structures also emerge in a CDM cosmology
(e.g., Cen et al.\markcite{Cen94} 1994).
Thus, the pancakes are thought to be a ubiquitous feature
in gravitational instability scenarios.
Here, we consider pancakes in the plane-parallel symmetry.
Also, previous 1D/2D/3D simulations on the pancake collapse 
in a dark matter-dominated universe
show that the pancake is finally dominated by baryons after the caustics
(Shapiro, Struck-Marcell, \& Melott \markcite{SSM83}1983; 
Bond et al.\markcite{BCSW84}1984; Shapiro \& Struck-Marcell \markcite{SS85}1985;
Yuan, Centrella, \& Norman \markcite{YCN91}1991; Umemura \markcite{Ume93}1993;
Cen et al. \markcite{Cen94}1994; 
Miralda-Escude \& Rees \markcite{MR94}1994; 
Hernquist et al.\markcite{HKWM95} 1995; 
Zhang, Anninos, \& Norman \markcite{ZAN95}1995; 
Anninos \& Norman \markcite{AN96}1996).
The presence of dark matter increases the shock velocity of the
falling matter into the caustics. 
According to previous analyses (e.g. Shapiro \& Kang \markcite{SK87}1987), such
shocks change the thermal evolution in the absence of a UV background
when the temperature exceeds $10^4$ K since the temperature controls
H$_2$ fraction in the postshock regions.
However, under UV background radiation, 
the difference of shock velocity will not alter the results significantly,
because the sheets are initially ionized by UV radiation 
and therefore quickly heated up to $\sim 10^4$K.
Thus, the thermal evolution of shocked region will be  
rather similar for any infall velocity, 
regardless of the shock temperature.
Thus, in order to highlight the relevant physics of thermal
processes, we deliberately exclude the dark matter contribution.
Nonetheless, some dynamical effects are anticipated from dark matter,
e.g. Jeans instability. They are discussed later in the light of the
present results.


In section 2, we describe the basic equations and initial
conditions. In section \ref{TDevolution}, the numerical 
results are presented. In section \ref{cond}, we make the physical 
interpretation of numerical results and 
present the condition on the cooling of a
collapsing pancake. Section \ref{implication}
is devoted to the implications for galaxy formation
under UV background radiation. In the last section, we summarize
the results.

\section{Formulation}

\subsection{Basic Equations}
\label{basic}

In this section, we give the basic equations
for hydrodynamical calculations. We assume the plane-parallel symmetry
throughout this paper.
A set of the ordinary hydrodynamical equations is given as below.
\begin{eqnarray}
\frac{d \left(1/\rho\right)}
{d t} - \frac{\partial u}{\partial m} 
& = & 0, \label{eqn:mcons} \\
\frac{d u}{d t}+\frac{\partial p}{\partial m} 
&=& g, \label{eqn:pcons}\\
\frac{d E}{d t}
+\frac{\partial \left( up \right)}{\partial m}
& = & ug - \Lambda(\rho,T,y_i), \label{eqn:econs}
\end{eqnarray}
where $\rho, p, u$, and $z$ denote
the density, the thermal pressure, the velocity,
and the length measured from the mid-plane, respectively, and
$E$ is the total energy per unit mass,
\begin{eqnarray}
E &=& \frac{p}{\rho \left( \gamma-1 \right)}+\frac{u^2}{2},
\end{eqnarray}
$m(z)$ is the column density,
\begin{eqnarray}
m(z) &\equiv& \int_0^z \rho(z) dz, \label{eqn:edef}
\end{eqnarray}
and $g$ is the gravitational acceleration,
\begin{eqnarray}
g &\equiv & 2\pi G m(z). \label{eqn:gdef}
\end{eqnarray}
The symbol $\Lambda(\rho,T,y_i)$ denotes
the cooling rate minus heating rate
per unit mass due to the radiative and chemical
processes of {\it i}-th species.
We take H$_2$ cooling rate from Hollenbach \& Mckee \markcite{HM79}
(1979), and the other atomic cooling rates are the same as those in
Shapiro \& Kang \markcite{SK87}(1987), 
except the cooling rate concerning on helium.
Photoheating rate is calculated by taking into account the radiation transfer 
effects. The formulation is given in Appendix \ref{imgamma}.
The equation of state is assumed as $p=\rho kT /\mu_M$, where $k$ denote
the Boltzman constant, and $\mu_M$ is the mean molecular weight of the
five species discussed below in this section. 
{The hydrodynamical equations are
solved by the Piecewise Parabolic Method (PPM) described in Colella \&
Woodward \markcite{CW84}(1984). 
PPM is one of the accurate methods to resolve the strong 
shock front. We take typically 800 spatial Lagrange grids.
The hydrodynamical routine is tested by the Sod's problem. This code
resolves the shock front typically by a few mesh. }

In the equations (\ref{eqn:mcons}), (\ref{eqn:pcons}), and
(\ref{eqn:econs}), the cosmic expansion is neglected.
It is because we have
considered the massive clouds far beyond the Jeans scale ($\lambda_J
\sim 10[(1+z)/10]^{-1.5}{\rm kpc}$) and 
also postulated that the initial stage is close to the maximum
expansion for a density fluctuation.
The effects of the cosmic expansion could be important 
for a density perturbation near the Jeans scale, since
the temperature is raised up to $\sim 10^4$ K
by UV background radiation in spite of the expansion
and consequently the dynamics could be significantly 
influenced by the UV heating 
(e.g. Umemura \& Ikeuchi \markcite{UI84}1984). 
However, far beyond the Jeans scale, 
the thermal pressure of gas around $10^4$ K
anyway does not affect the dynamical evolution until the maximum expansion
as shown by previous calculations (e.g. Thoul \& Weinberg \markcite{TW96}1996). 
Therefore, we attempt to pursue the history 
after the maximum expansion.

The hydrodynamical equations are coupled with 
the non-equilibrium rate equations for chemical reactions:
\begin{eqnarray}
\frac{d y_i}{d t} = \sum_{j} k_j y_j
+n \sum_{k,l} k_{kl}y_k y_l
+n^2 \sum_{m,n,s} k_{mns}y_m y_n y_s, \label{eqn:chemgeneral}
\end{eqnarray}
where $y_i$ is the fraction of {\it i}-th species,
$y_i \equiv n_i/n$, 
with $n$ being the number density of hydrogen nuclei, and $k$'s 
are the coefficients of reaction rates (see Table \ref{tab:reactions}).
In the present calculations, we take into account six species e, H, H$^+$, 
H$^-$, H$_2$, and H$_2^+$. 
Here, we neglect helium. This is because the helium lines are 
not an effective coolant for lower temperature ($T \lsim 10^4$ K) 
in which we are here especially interested.
Nonetheless, some of helium lines including recombination lines 
may affect the ionization structure, 
because they are energetic enough to ionize HI.
However, according to Osterborck
\markcite{Ost89}(1989), the difference between 
the ionization structure of pure
hydrogen and hydrogen + helium gas for 40000 K black body radiation 
is small. In the case of 
the power-law spectrum of UV background radiation which we 
consider, the difference could be larger, because the higher
energy photons contribute more to the ionization of helium 
compared to the case of black body radiation. 
In fact, in the recent calculation by Abel \& Haehnelt
\markcite{AH99}(1999), in which the radiation transfer effects of helium
recombination photons are taken into account, the temperature of the
clouds could differ by a factor of 2.
Although we have dismissed helium to focus upon elucidating 
the key physics, we should keep in mind that helium could be 
important in highly quantitative arguments.

The photoionization and heating processes 
due to the external UV radiation 
are pursued by solving the frequency-dependent 
radiative transfer equation for hydrogen, 
\begin{equation}
 \mu \frac{dI_\nu}{dz}
 = - \kappa_\nu \, I_\nu + \eta_\nu,
\label{eq:RTslab}
\end{equation}
where $\mu$, $\kappa_\nu$, and $\eta_\nu$ denote the direction cosine,
the absorption coefficient, and the emissivity, respectively.
The frequency-integration in the transfer equation can be done
analytically for the continuum, while the recombination line near 
the Lyman limit is separately treated by solving transfer equation
with a source term which comes from the recombination. 
The details are described in Appendix \ref{imgamma}.
{ Equation (\ref{eq:RTslab}) assumes steadiness of the radiation fields.
When a neutral cloud is exposed to background ionizing radiation,
ionized regions spread inward. In the ionized regions, the radiation
propagates basically at the light speed, which is typically $10^3$ times 
larger than the hydrodynamical velocity of the system. 
Thus, as for the radiation fields, it is sufficient to
solve the steady radiative transfer equation. But, the ionization
front propagates with a different speed, which is determined
by the balance between the number of neutral atoms flowing 
through the front per second and the corresponding number of
ionizing photons reaching the front. Resultantly, 
the speed of ionization front is much lower than the light speed, 
typically 500 km s$^{-1}$ $(n/10^{-2} {\rm cm^{-3}})^{-1}$
for UV background of $10^{-21}{\rm erg\; s^{-1} cm^{-2} str^{-1} Hz^{-1}}$.
In order to capture the ionization front propagation, 
we take the following numerical procedures. 
First, the steady radiative transfer equation (\ref{eq:RTslab}) is solved, 
where typically 800 spatial grids are used, which number is equivalent to 
the number of Lagrange meshes assigned for hydrodynamics, 
and 40 angular grids are adopted. 
We have checked that the resultant accuracy is not significantly changed 
if we use the larger number of angular grids.
The integration of equation (\ref{eq:RTslab}) is performed 
with the summation of the exact formal solution in every grid 
(Stone, Miharas, \& Norman \markcite{SMN92}1992).
This method is accurate, 
even if the optical depth of each grid can exceeds unity.
Also, the validity of this method is enhanced by the fact that 
the ionization front is not sharply edged, but fairly dull 
for a power law of UV radiation 
(Tajiri \& Umemura \markcite{TU98}1998). 
For the photons which cause H$_2$ dissociation
through the Solomon process (e.g. HRL\markcite{HRL96}1996), 
we employ the self-shielding function given by
Draine \& Bertoldi \markcite{DB96} (1996),
instead of directly solving the radiation transfer equation.
Next, the rate equations including photoionization are solved implicitly 
for a time step which is determined as bellow. 
Finally, the hydrodynamical equations with the thermal equation
is integrated explicitly.} 

{ The time step of the calculation $\Delta t$ is taken as follows: 
\begin{eqnarray}
\Delta t&=&
{\rm {min}} (0.99  t^{\rm hydro}, 
0.8  t^{\rm grav}, 
0.3  t^{\rm cool}, 
0.3  t^{\rm heat}).
\end{eqnarray}
Here $t^{\rm hydro}$ represents the hydrodynamical time, and
$0.99 t^{\rm hydro}$ is employed so that the Courant condition
is satisfied. 
 $t^{\rm grav}$ denotes the time scale of gravity, which is
defined as $t^{\rm grav}\equiv \sqrt{p/\gamma \rho}/|g|$, 
where $p$, $\gamma$,  $\rho$, and $g$ 
represents the pressure, the adiabatic index, the density, 
and the gravitational acceleration, respectively.
$t^{\rm cool}$ and $t^{\rm heat}$ represent 
local cooling time, and  heating time scale, respectively.
With these procedures, we have checked the propagation speed of the
ionization front defined by the point $y_{\rm HI}=0.1$. In all of our
simulations, the I-front propagation speeds are less than 1 percent of
the speed of light. Thus, the assumption of the steady radiation
transfer is valid. 
}

\subsection{Initial Conditions}

We assume the initial density distributions to have a cosine profile;
\begin{equation}
n_{\rm ini}(z)=\bar{n}_{\rm ini}
\left[1+\varepsilon \cos \left(\pi z/\lambda\right )\right],
\end{equation}
where $\bar{n}_{\rm ini}$ and $\lambda$ are the initial mean 
number density and the thickness of the sheet (See Fig.1), 
and $\varepsilon$ denotes
the density contrast, which is set to be 0.5 throughout this paper.
The ranges of $\bar{n}_{\rm ini}$ and $\lambda$ 
which we consider are respectively $10^{-6} {\rm cm^{-3}} \le
\bar{n}_{\rm ini} \le 5\times 10^{-2}{\rm cm^{-3}} $ and $0.3 {\rm
kpc} \le \lambda \le 3 {\rm Mpc}$.
%
The initial velocity is null for every mass layer,
because the initial stage is implicitly assumed to be
close to the maximum expansion stage of a density fluctuation.
In general, an overdense region at the maximum expansion is surrounded
by an underdense region which is still expanding.
Hence, if we introduce the initial velocity distributions 
more realistically, the further mass may accrete from the underdense 
region after the pancake collapse. However,
the envelope does not seem to contribute to the self-shielding
because of its low density, and thus the self-shielding is 
determined by the column density of the first collapsed sheet. 
Hence, we neglect the surrounding expanding 
underdense region, and model only the overdense region.
Moreover, the velocity gradient may exist in a collapsing
overdense region, and then leads to the delay of mass accretion 
from outer regions. The degree of self-shielding is determined 
by the total recombination number per unit time ($N_{\rm rec}$) in the volume 
against the UV photon number per unit time ($N_{\rm UV}$) from the boundary.
If the envelope has a density distribution in proportion to $r^x$,
$N_{\rm rec}$ is proportional to $r^{2x+1}$, whereas $N_{\rm UV}$ is constant,
as far as the envelope undergoes nearly sheet-like collapse. 
Thus, an envelope with a distribution steeper than $r^{-1/2}$
would not contribute to the self-shielding.
In other words, the inner regions which collapse 
nearly simultaneously contribute mainly to the self-shielding.
Although there are such qualitative expectations, we cannot assess 
accurately the effects by the initial velocity gradient and the
cosmological expansion, because we have not included them.
They should be quantitatively investigated in the future analyses.

The initial temperature and chemical composition 
are given respectively 
by the thermal equilibrium and by the ionization equilibrium.
Also, the fractions of H$_2$, H$^-$ and H$_2^+$ are 
initially determined by the chemical equilibrium.
When determining the initial equilibrium state,
H$_2$ cooling rate is ignored, because
we set the initial condition so that
the UV heating should overwhelm the H$_2$ cooling.
It will be shown below that this condition can be satisfied 
even if the cloud is almost neutral.

The incident radiation intensity, $I_\nu^{\rm in}$, is assumed to have 
a power low spectrum as 
\begin{equation}
I_\nu^{\rm in}=I_{21}(\nu/\nu_L)^{-\alpha} 10^{-21}{\rm erg\; s^{-1} cm^{-2} str^{-1} Hz^{-1}} , 
\end{equation}
where $\nu_L$ is the frequency at the Lyman limit.
The observations require
the diffuse UV radiation to be at a level of $I_{21}=10^{\pm 0.5}$
at $z=1.7-4.1$ (Bajtlik, Duncan, \& Ostriker \markcite{BDO88}1988; 
Giallongo et al. \markcite{GCD96}1996).
At higher redshifts, we have no indication for $I_{21}$.
Here, we assume $I_{21}=0.1$, and the scaling to $I_\nu^{\rm in}$ is argued. 
As for $\alpha$, two typical cases are considered, i.e. $\alpha=1$ or 5, 
which are the same choice as Thoul \& Weinberg \markcite{TW96}(1996). 
The former represents a quasar-like spectrum, 
and the latter resembles a spectrum of star-forming galaxies
around the Lyman limit frequency.

\section{Thermal and Dynamical Evolution}
\label{TDevolution}

In this section, we show the characteristic behaviors of the
thermal/dynamical evolution resulting from numerical runs, 
and elucidate the dependence of the final states 
upon the initial states.

\subsection{Hard Spectrum Case: $\alpha=1$}

Figures 2 and 3 
show the time evolution of spatial distributions of various 
physical quantities in the case of $\alpha=1$. 
Figure 2 represents the case
with $\lambda=100$ kpc and $\bar{n}_{\rm ini}= 10^{-2}$ cm$^{-3}$. 
In this figure, each horizontal axis represents the Lagrange coordinate, normalized by 
the half total surface density, $m_0$.
The thick line shows the distribution at
$t=1.28/\sqrt{4\pi G \bar{\rho}_{\rm ini}}$,
where $\bar{\rho}_{\rm ini}=m_p \bar{n}_{\rm ini}$ with
$m_p$ being the proton mass. The calculations are terminated there.
Figure 2 shows that in the course of the collapse, 
a thin cold sheet quickly forms due to the efficient molecular cooling,
which is confined by a shock-heated high temperature layer.
The lower panels in Figure 2 show 
the corresponding HI and H$_2$ distributions.
Only an outer envelope ($m/m_0 \gsim 0.9$) is photoionized and
the rest parts of the sheet are self-shielded against 
UV radiation. As a result, the sheet is mostly neutral 
throughout the dynamical evolution, excepting the
collisionally ionized postshock regions.
In the early phase, UV heating exceeds the H$_2$ line cooling 
even in the almost neutral regions.
The H$_2$ cooling becomes efficient as the collapse proceeds
to a further degree, and finally 
the temperature drops down to $\sim 100$ K.
Then, the collisional dissociation of hydrogen
molecules becomes ineffective, and thus 
the fraction of hydrogen molecules at $m=0$
continues to increase.

Figure 3 represents the case with lower density
and larger size, i.e., 
$\bar{n}_{\rm ini}=5\times 10^{-5}{\rm cm^{-3}}$ and $\lambda=1$ Mpc. 
The sheet also collapses because the initial thickness is larger than the
initial Jeans length. 
However, the central temperature does not drop below $\sim 10^4$ K,
and the resultant H$_2$ fraction is too low to overwhelm the UV heating. 
Why the evolution results in 
the different final states from the case shown in Figure 2 is basically understood by
the difference of the degree of self-shielding against the external UV radiation.
In the previous case, the most regions are so strongly shielded from 
the external UV that 
abundant H$_2$ form and the cooling rate due to H$_2$ becomes predominant. 
Contrastively, in the present case, the external UV permeates the inner regions
to a considerable degree, so that UV heating overwhelms
H$_2$ cooling everywhere until the collapse stops due to the thermal
pressure.

\subsection{Soft Spectrum Case: $\alpha=5$}

Figure 4 shows the case of a softer spectrum, $\alpha=5$.
The other initial parameters are the same as Figure 3. 
Compared to the previous hard spectrum cases, 
the inner parts of the sheet are quickly self-shielded 
as the collapse proceeds. This is due to the fact that
there are fewer high energy photons which penetrate the deep inside 
the sheet owing to the smaller ionization cross-section 
(see e.g. Tajiri \& Umemura \markcite{TU98}1998).
As a result, the photoionization rate as well as the UV
heating rate damps strongly in the inner regions, so that
a cold layer ($T< 10^4$ K) due to H$_2$ cooling emerges
eventually. 
The self-shielded regions grow thicker and the the fraction of H$_2$ 
goes up to a level of $10^{-3}$ similar to the case of Figure 2.

\subsection{Summary of the Dependence on Initial States}

In Figure 5, the final physical states of the sheets for
$\alpha=1$ at $t=1.28/\sqrt{4\pi G\rho_{\rm ini}}$ 
are summarized on the initial density-to-size diagram. 
Depending upon the initial thickness $\lambda$ 
and mean density $\bar{n}_{\rm ini}$, the sheets evolve in different ways. 
Filled circles in Figure 5 denote the initial conditions
on which the clouds eventually cool below $5\times 10^3$ K 
during the collapse. Open circles denote the ones on which the clouds 
cannot cool ($T >  5\times 10^3$ K).
The boundary of the evolutionary bifurcation
is well fitted by
a simple power low relation as
\begin{eqnarray}
\lambda_{\rm cool} = 1.1 {\rm Mpc}
\left(\frac{\bar{n}_{\rm ini}}{1.0 \times 10^{-4} {\rm cm^{-3}}}\right)^{-0.8}.
\end{eqnarray}
The small open squares, open triangles, and filled triangles 
in Figure  5 trace the
evolutionary paths for three typical parameters. 
In the paths, the time-dependent thickness ($H$) 
is defined by the thickness within
which 80\% of the total mass of the sheet is contained 
and the time-dependent mean density ($\bar{n}$) 
is defined by the spatially averaged density over $H$;
\begin{eqnarray}
H&\equiv& 2 \int_0^{0.9m_0}\frac{1}{\rho}dm, \\
\bar{n} &\equiv& \frac{2}{H} \int_0^H n dz.
\end{eqnarray}
The open squares and open triangles are respectively
the results for the initial parameters of
$(\lambda,\bar{n}_{\rm ini})
=(1{\rm Mpc}, 5\times 10^{-5} {\rm cm^{-3}})$ 
and $(100{\rm kpc}, 1\times 10^{-3}{\rm cm^{-3}})$.
For such initial conditions, the sheet shrinks due to 
Jeans instability [$\lambda > \lambda_J(T=10^4 {\rm K})$].
However, the collapse ceases by the thermal pressure before it
intersects the critical line marked by $\lambda_{\rm shield}$, above which
the sheet can be self-shielded and cool down
due to formed H$_2$. 
(The further details on $\lambda_{\rm shield}$ 
are discussed in the next section.)
On the other hand, the filled triangles, corresponding to the run 
with $(\lambda,\bar{n}_{\rm ini})
= (100 {\rm kpc}, 1\times 10^{-2} {\rm cm^{-3}})$, come into the regions
$\lambda > \lambda_{\rm shield}$ before the thermal pressure prevents the sheet
from collapsing. 
Consequently, this sheet is shielded and cools below $T\sim 5\times 10^3 K$.

Figure 6 shows the dependence on the initial states
for $\alpha=5$. 
Because of the smaller number of runs, 
it is somewhat hard to divide clearly the parameter space.
If we dare to draw the boundary, it is proportional 
to $\bar{n}_{\rm ini}^{-0.66}$, which is numerically
\begin{eqnarray}
\lambda_{\rm cool}=6.0\times 10^2 {\rm kpc}
\left(\frac{\bar{n}_{\rm ini}}{1.0\times 10^{-4}
{\rm cm^{-3}}}\right)^{-0.66}.
\end{eqnarray}
This criterion is slightly different from the previous case,
in the sense
that $\lambda_{\rm cool}$ for $\alpha=5$ does not cross 
the intersection point of $\lambda_J(T=10^4{\rm K})=\lambda_{\rm shield}$, 
although $\lambda_{\rm cool}$ for $\alpha=1$ does.

\section{Physical Conditions for H$_2$ Cooling under UV background}
\label{cond}

In this section, we attempt to understand the underlying physics of the
present numerical results.
To begin with, we introduce a scale, $\lambda_{\rm shield}$, which
characterizes the relative efficiency of H$_2$ cooling to UV heating.
Then, coupled with some physical arguments,
we try to estimate the boundary of the bifurcation, 
$\lambda_{\rm cool}$, which has been derived by the present numerical
calculations. 
\par
As a measure of the penetration of UV photons into the
cloud, a shielding length, $\lambda_{\rm shield}$, 
is defined by the balance between the UV heating rate 
and the H$_2$ cooling rate, which are respectively 
\begin{eqnarray}
{\rm Heating\; Rate} &=& n(z)y_{\rm HI}
\Gamma_{\rm HI}\left(\tau_{\nu_L,v}\left( z \right)\right),
\label{eq:hrate}\\
{\rm Cooling\; Rate} &=& n(z)^2\Lambda_{\rm H_2}
(y_{H_2}(z),T(z))\label{eq:crate},
\end{eqnarray}
where $\Gamma_{\rm HI}$ is the UV heating rate per particle and
$\Lambda_{\rm H_2}$ denotes the H$_2$ cooling function.
As shown in Appendix \ref{imgamma} [eq.(\ref{eq:asymheat})], 
in an optically thick regime, 
the UV heating rate is expressed by a power law of the optical depth
at the Lyman limit, 
not by an exponential law. This is again because higher energy photons
have the smaller ionization cross-section.
The optical depth at the Lyman limit measured from the outer
boundary ($z=z_{\rm out}$) is
\begin{eqnarray}
\tau_{\nu_L,v}(z)&\equiv&\int_z^{z_{\rm out}}dz 
~n y_{\rm HI}\sigma_{\nu_L},
\end{eqnarray}
where $\sigma_{\nu}$ is the ionization
cross-section.
If the slab is almost neutral,
\begin{eqnarray}
\tau_{\nu_L,v}(0)&\simeq&\int_0^{z_{\rm out}} dz 
~n \sigma_{\nu_L},\\
&=& \bar{n}_{\rm ini} \sigma_{\nu_{L}} \lambda /2. 
\label{eq:odepth}
\end{eqnarray}
Equating (\ref{eq:hrate}) with (\ref{eq:crate})
at the midplane of the slab ($z=0$), and also using 
equations (\ref{eq:odepth}) and (\ref{eq:asymheat}),
we can assess $\lambda_{\rm shield}$ as 
\begin{equation}
\lambda_{\rm shield}\simeq
\frac{2}{\bar{n}_{\rm ini}\sigma_{\nu_L}}
\left(\frac{2\pi I_{\nu_L}^{\rm in}\sigma_{\nu_L}\nu_L}
{3\bar{n}_{\rm ini}\Lambda_{\rm H_2}}
\frac{\Gamma\left(\beta \right)}{1+\beta}
\right)^{\frac{1}{\beta}}\propto \bar{n}_{\rm ini}^{-1-1/\beta}
\label{eq:lambdac}
\end{equation}
where $\beta\equiv 1+\left(\alpha-1\right)/3$.
The $\lambda_{\rm shield}$ is plotted in 
Figures 5 and 6 
by long-dashed lines.
The cooling rate has been assumed to be 
$\Lambda_{\rm H_2}=10^{-26}{\rm cm^3s^{-1}}$, 
which is a maximal value of H$_2$ cooling 
for the primordial gas that is once heated up to 
$T\gsim 10^4 {\rm K}$ 
(Shapiro \& Kang \markcite{SK87}1987; Susa et al. \markcite{SUNY98}1998).
\par
The obtained $\lambda_{\rm shield}$ represents the degree of
self-shielding against the external UV at the initial stages:
(1) For $\lambda > \lambda_{\rm shield}(\bar{n}_{\rm ini})$, 
the sheet is promptly self-shielded enough to be cooled by the H$_2$. 
(2) For $\lambda < \lambda_{\rm shield}(\bar{n}_{\rm ini})$, the sheet 
is not self-shielded initially, so that the gas is heated up to 
$T\gsim 10^4$ K due to predominant UV heating.
A similar argument with $\lambda_{\rm shield}$ is also 
applicable to each dynamical stage. 
In the case, we should just replace the density $\bar{n}_{\rm ini}$ 
by $\bar{n}$, and $\lambda$ by $H$, both of 
which are defined in section \ref{TDevolution}. 
Then, we interpret $\lambda_{\rm shield}$ as the
boundary beyond which the sheet is quickly self-shielded
and consequently cools down owing to the efficient formation of H$_2$.
In fact, the clouds denoted by filled circles
in Figure 5 satisfy the condition 
$H > \lambda_{\rm shield}$ before the collapse is abruptly 
decelerated by approaching $\lambda_J(T=10^4{\rm K})$.
Eventually, the clouds cool down below $5\times 10^3$ K.
On the other hand, the clouds denoted by open circles
above $\lambda_J$  in Figure 5 result in
warm sheets of $\sim 10^4$ K, because the clouds first meet  
$\lambda_J(T=10^4{\rm K})$ before intersect $\lambda_{\rm shield}$.
If we assume that the sheet clouds evolve keeping 
the relation $\bar{n} H = {\rm const.}$ as anticipated 
in the ideal sheet collapse, 
the boundary of the bifurcation in the space $(\bar{n}_{\rm ini},\lambda)$ 
is expected to be a line which satisfies 
$\bar{n}_{\rm ini}\lambda = {\rm const.}$ and meets the intersection 
point of $\lambda_{\rm shield}=\lambda_J$. 
This prediction, $\lambda_{\rm cool} \propto \bar{n}_{\rm ini}^{-1}$,
is close to the numerically obtained boundary,
$\lambda_{\rm cool} \propto \bar{n}_{\rm ini}^{-0.8}$.
The small difference of the dependence may come from the fact that
the simulations contain the spatial structure of density and temperature,
and therefore partial self-shielding.

We remark that the Jeans length $\lambda_J$ is maximally scaled by
$\sqrt{\Omega_{\rm B}/\Omega_0}$ in the
presence of dark matter,
where $\Omega_{\rm B}$ is the baryon density parameter
and $\Omega_0$ is the total density parameter. 
However, the baryonic component dominates the gravity of
the sheets in the final phase of the collapse, although dark matter
component does initially. Thus, what actually happens is the intermediate
of these two extreme cases. Although calculations that include dark matter are necessary to evaluate the effect quantitatively, we leave them elsewhere. 

\section{Implications for Galaxy Formation}
\label{implication}

Based upon the present numerical results, we consider 
the context of the galaxy formation under UV background
radiation. 
First of all, the cooling by atomic processes of primordial gas 
is essential for the formation of H$_2$ molecules which control
the star formation in primordial objects. The elaborate analyses
by Rees \& Ostriker \markcite{RO77}(1977) and 
also by Blumenthal et al. \markcite{BFPR84}(1984)
show that the cooling mass is roughly constant almost regardless of
the virial temperature and density, which is $\sim 10^{12}M_\odot$.
If there is UV background radiation, the cooling mass could alter
because the line cooling by H and He at $\lsim 10^5$K is 
seriously reduced (e.g. Thoul \& Weinberg \markcite{TW96}1996). The mass, however,
is basically determined by the cooling mechanisms at 
$\gsim 10^5$K. At such temperature, 
the cooling in photoionized gas is dominated by collisional ionization, 
radiative recombination, and thermal bremsstrahlung, 
and possibly the Compton cooling at high redshifts.
These mechanisms may potentially cool the clouds 
with $\sim 10^{12}M_\odot$ down to several $10^4$K.  
In fact, it is shown by 
numerical calculations that the clouds with $\sim 10^{12}M_\odot$
can cool and collapse under UV background (Umemura \& Ikeuchi \markcite{UI84}1984).
Thus, we set here the upper mass of primordial galaxies 
to be $10^{12}M_\odot$. Then,
the evolution of pregalactic clouds under UV background
radiation is discriminated into four categories; (a) promptly
self-shielded clouds, (b) starburst pancakes, (c) retarded
star-forming galaxies, 
and (d) expanding clouds. 
In Figure 7, the parameter regions corresponding
to the four categories are shown for $\alpha=1$, with
lines of equal mass, say, $10^{8}M_\odot$ and
$10^{12}M_\odot$. 
The upper abscissa is the collapse redshifts, $z_c$, 
if the initial stage is assumed to be at the maximum expansion
of a spherical top-hat density fluctuation.
$\Omega_0=1$, $h=0.5$, and $\Omega_{\rm B} h^2 =0.02$ are assumed in order
to interpret the maximal expansion density into the collapse redshift $z_c$,
where $h$ is the present Hubble constant in units of ${\rm 100km~s^{-1}Mpc^{-1}}$.
In this figure, 1$\sigma$, 2$\sigma$, and 3$\sigma$
density fluctuations expected in a standard CDM are also shown.
Some further details for each category are discussed in the following,
restricting ourselves to the case $\alpha=1$.
\subsection{Promptly Shielded Clouds}
The evolution of promptly self-shielded clouds above
$\lambda_{\rm shield}$ is virtually equivalent to the evolution
under no UV background radiation. The collapse redshifts of such clouds 
are expected to be $z_c \gsim 10$.
The formation of primordial objects under no UV background
has been hitherto extensively
studied by numerous authors 
(Matsuda, Sato, \& Takeda \markcite{MST65} 1965; Yoneyama \markcite{YO72}1972;
Hutchins \markcite{HU76}1976; Palla, Salpeter, \& Stahler \markcite{PSS83}1983;
Susa, Uehara, \& Nishi \markcite{SUN96} 1996; 
Annonis \& Norman \markcite{AN96} 1996; 
Tegmark et al. \markcite{TSR97} 1997).
As a result, we can expect the first generation of objects in the mass 
range of $10^{6}M_\odot$ at $z_c=15$ down toward
$10^{3}M_\odot$ at $z_c \gg 200$ (Tegmark et al. \markcite{TSR97} 1997).

\subsection{Initial Starbursts in Pancakes}

The region (b) in Figure 7 results in
cold pancakes due to H$_2$ cooling, in which initial starbursts 
may take place.
The instability of a 
shock-compressed layer has been discussed by several authors (Elmegreen \&
Elmegreen \markcite{EE78}1978; Vishniac\markcite{Vis83} 1983; Lubow \& Pringle \markcite{LP93}1993; Whitworth
et al. \markcite{WBCDT94}1994; Yamada \& Nishi \markcite{YN98}1998). According to Elmegreen \& Elmegreen \markcite{EE78}(1978), 
the fastest growing modes have the size of the sheet thickness and 
the fragmentation timescale is
$t_{\rm frag}=(-4\pi\rho(0)w^2)^{-1/2},$
where $w^2\sim -0.14$ in the high external pressure limit.
We estimate the line mass $l_0$ at fragmentation by a condition
$t_{\rm frag}=t_{dyn}$, where $t_{dyn}=\rho(0)/\dot{\rho}(0)$.
Under an assumption that cold ($T\lsim 10^3$ K) sheets fragment into filaments, 
the resultant line mass is tabulated in Table \ref{tab:frag}
with other properties, where
$f$ is the ratio of $l_0$ to the critical line density 
$l_c \equiv 2kT_0/m_p G$, and
$x_s$ and $x_c$ denote the mass fraction of the shocked matter and
cooled matter, respectively.
It is noted that $f$ is smaller than unity for the fastest growing modes.
Hence, for the fragmentation,
it is necessary for other growing modes 
to accumulate further mass so that the filaments would be super-critical. 
The super-critical filaments would eventually bear massive
stars (Nakamura \& Umemura \markcite{NU99} 1999). 

We can estimate the minimum mass of the starburst pancakes by
$M_{\rm SB} \equiv \bar{\rho}_{\rm ini} \lambda_{\rm cool}^3$.
Again, if the initial stage is at the maximum expansion, then
\begin{equation}
M_{\rm SB} = 2.2\times 10^{11} M_\odot \left(\frac{1+z_c}{5}\right)^{-4.2} 
\left(\frac{\Omega_{\rm B} h^2}{0.02}\right)^{-1.4}.
\end{equation}
The collapse epochs of the clouds in region (b) 
range from $z_c \simeq 3$ to $z_c \simeq 10$.
The clouds may undergo further shrinking after the sheet-like collapse,
because the rotation barrier is smaller by $\Lambda_{\rm spin}^2$ than
the maximum expansion size, where $\Lambda_{\rm spin}$ is 
the dimensionless spin parameter which is peaked around
0.05 (Heavens \& Peacock \markcite{HP88}1988). Consequently,
the violent relaxation would take place in a collisionless fashion.
Hence, the starburst pancakes highly possibly lead to 
the dissipationless galaxy formation.
It is worthy noting that $M_{\rm SB}$ lies between 
1$\sigma$ and 2$\sigma$ fluctuations in a standard CDM scenario.
Thus, they could form a relatively clustered population
of galaxies.

\subsection{Retarded Galaxy Formation}

The clouds with the smaller mass than $M_{\rm SB}$ 
cannot cool below $10^4$K at the first pancake collapse, 
but they might pass $\lambda_{\rm shield}$ in the course of
shrinking down to the rotation barrier, so that the clouds
are self-shielded against external UV radiation to cool down
by H$_2$ molecules.  
Assuming $\Lambda_{\rm spin}=0.05$, the marginal size, 
$\lambda_{\rm rot}$, above which the clouds
can pass $\lambda_{\rm shield}$ is much smaller than size of 
the Jeans unstable clouds for $z_c \ge 0$. Thus, no cloud can be
stopped by the rotation barrier before the cloud is self-shielded.
In the clouds in the region (c),
the star formation is likely to be retarded until the pancake
disk shrinks considerably. Therefore, the region (c) tends to result 
in the so-called dissipational galaxy formation 
(e.g. Larson \markcite{Lar76}1976; Carlberg \markcite{Car85}1985; 
Katz \& Gunn\markcite{KG91} 1991).
The collapse epochs are relatively later ($0\lsim z_c \lsim 4$),
compared to the region (b), and the region (c) is corresponding
to CDM fluctuations lower than 1$\sigma$.
Thus, it is likely that the region (c) leads to a less
clustered population of galaxies.
Futhermore, for low redshifts ($z\lsim 1$), the intensity of UV background
is no longer at a level of $I_{21}=0.1$, but could be smaller
by two orders of magnitude 
(Maloney \markcite{Mal93}1993; Henry \& Murthy\markcite{HM93} 1993; Dove
\& Shull \markcite{DS94}1994).
Since $\lambda_{\rm shield} \propto I_{21}$ for $\alpha=1$ 
as seen in (\ref{eq:lambdac}), the clouds with $z\lsim 1$ 
in the region (c) might cool quickly in the course of
shrinking down to the rotation barrier. 

\subsection{Expanding Clouds}

In the region (d), the clouds are no longer gravitationally bound
because of the enhancement of thermal pressure due to UV heating.
They are relatively low mass systems.
They could be Lyman alpha absorption systems which are seen in QSO
spectra (Umemura \& Ikeuchi \markcite{UI84}1984, 1985; Bond, Szalay, \& Silk \markcite{BSS88}1988).


\section{CONCLUSIONS}
\label{conc}

We have numerically explored the thermal and dynamical evolution
of pregalactic clouds under UV background radiation.
The plane-parallel collapse of primordial gas clouds
has been pursued, including chemical reactions with respect to
hydrogen molecules as well as atomic hydrogen. Also,
the radiation transfer for the ionizing photons has been
properly treated.
As a result, it is found that the cloud evolution under UV background
branches off into four categories in the initial parameter space
of density and size. They are (a) promptly
self-shielded clouds, which evolve into dense objects 
with $\lsim 10^6M_\odot$ at collapse redshifts ($z_c$) greater than 10, 
(b) starburst pancake clouds with the mass higher than
$M_{\rm SB} = 2.2\times 10^{11} M_\odot \left[(1+z_c)/5\right]^{-4.2}$, 
which lead to the dissipationless galaxy 
formation at $3\lsim z_c \lsim 10$, (c) retarded
star-forming galaxies with the mass lower than $M_{\rm SB}$,
which undergo star formation in the course of
shrinking down to the rotation barrier at $0\lsim z_c \lsim 4$, 
consequently leading to the dissipational galaxy formation, 
and (d) less massive expanding clouds,
which could be detected as Lyman alpha absorbers
in QSO spectra.
If we assume a standard CDM cosmology, 
density fluctuations of 1$\sigma$-2$\sigma$ 
coincides with $M_{\rm SB}$. That is, fluctuations higher 
than 2$\sigma$ result in the dissipationless formation of 
massive galaxies at $3\lsim z_c \lsim 10$, and
eventually constitute a clustered population.
From a further realistic point of view, there must be
local subgalactic clumps growing in a pregalactic cloud
in a bottom-up scenario like a CDM cosmology.
They are likely to be self-shielded earlier than 
uniform components. Thus, $M_{\rm SB}$ can be considered
as a measure of determining the bulge-disk ratios (B/D) of formed galaxies.
Above $M_{\rm SB}$, the B/D ratio should be larger, while 
the ratio becomes smaller below $M_{\rm SB}$.

\acknowledgments
We are very grateful to Steven N. Shore for reviewing the article 
and making valuable comments and criticism, 
and also for enormous assistance as a editor.
We also thank T. Nakamoto for continuous encouragement. 
We thank also R. Nishi and H. Uehara for useful discussions.
This work is supported in part by Research Fellowships of the Japan Society 
for the Promotion of Science for Young Scientists, No.2370 (HS), and
the Grants-in Aid of the
Ministry of Education, Science, Culture, and Sports, 09874055 (MU).
\appendix
\section{Frequency Integration for Photoionization and UV Heating Rate}
\label{imgamma}
In this appendix, a method to integrate the frequency 
dependence in radiation transfer equation is presented 
(see also Tajiri \& Umemura \markcite{TU98}1998).  
This method is applied to plane-parallel calculations in this paper. 
However, it is also potent for the 3D calculations on cosmic
ionization problem (Nakamoto, Umemura \& Susa \markcite{NUS99}1999). 
The frequency-dependent radiation transfer equation is 
given by a general form as,
\begin{equation}
 \mbox{\boldmath$n$} \cdot \nabla \, I_\nu
 = - \kappa_\nu \, I_\nu + \eta_\nu
 \label{eq:RT}
\end{equation}
where {\boldmath$n$} is the unit directional vector,
$\kappa_\nu$ and  $\eta_\nu$ denote the opacity and the emissivity,
respectively.
For the gas composed of pure hydrogen, the emissivity is almost 
null for $\nu\gsim
\nu_L+\Delta\nu_T$, where
$\Delta\nu_T$ denotes the width of the Lyman limit emission
resulting from radiative recombination;
\begin{equation}
\Delta\nu_T \equiv kT/h_{\rm P},
\end{equation}
where $T$, $k$ and $h_{\rm P}$ denote 
the gas temperature, and
Planck constant, respectively. 
Typically, $\Delta\nu_T/\nu_L \approx 0.1$ for $T=10^4$K.
As far as higher energy photons of $\nu\gsim \nu_L+\Delta\nu_T$
are concerned, the solution of equation (\ref{eq:RT}) is simply
\begin{eqnarray}
I_\nu = I_\nu^{\rm in}
\exp\left(-\tau_\nu\left(\omega\right)\right) \label{eq:dump},
\end{eqnarray} 
where $I_\nu^{\rm in}$ denotes the incident intensity at the boundary.
The optical depth $\tau_\nu(\omega)$ in the solid angle
$\omega$ is rewritten in terms of the optical depth
at Lyman limit, $\tau_{\nu_L}(\omega)$,
\begin{eqnarray}
\tau_\nu(\omega)=\tau_{\nu_L}(\omega)\left(\nu_L/\nu\right)^3, \label{eq:tau}
\end{eqnarray}
because the photoionization cross-section is 
$\sigma_{\nu}\simeq\sigma_{\nu_L}(\nu_L/\nu)^3$.
Using the equations (\ref{eq:dump}) and (\ref{eq:tau}), we obtain 
the photoionization rate coefficient for this energy range of photons, 
$k_{\rm HI}^{abs}$, in terms of an integration as
\begin{eqnarray}
k_{\rm HI}^{abs}& =& \int_{\nu_L+\Delta \nu_T }^\infty \int \frac{I_\nu}{h\nu} \sigma_{\nu} d\omega d\nu, \nonumber  \\
          & \simeq& \frac{I_{\nu_L}^{\rm in}\sigma_{\nu_L}}{h}\int
\frac{1}{3\tau_{\nu_L}(\omega)^{1+\alpha/3}}\gamma(1+\alpha/3,\tau_{\nu_L}\left(\omega\right))d\omega, \\
\;&\propto& \tau_{\nu_L,v}^{-1-\alpha/3}\;\;\;(\tau_{\nu_L,v}\gg 1 ),
\label{eq:ion1abs}
\end{eqnarray}
where we have assumed
$I_\nu^{\rm in}=I_{\nu_L}^{\rm in}(\nu/\nu_L)^{-\alpha}$
and $\gamma(a,b)$ denotes the incomplete gamma function. 
$\tau_{\nu_L,v}$ denotes the optical depth measured in
the vertical directions of a slab,
which is already introduced in section \ref{cond}.
It is noted that the integration in frequency space has been carried 
out just analytically with the incomplete gamma function.
In other words, we do not have
to solve the radiation transfer equation for $\nu \gsim \nu_L +
\Delta\nu_T$.

On the other hand, photons with $\nu_L \le \nu \le \nu_L + \Delta\nu$
are scattered to produce diffuse radiation.
For hydrogen, the scattering albedo is given by
$[\alpha_A(T)-\alpha_B(T)]/\alpha_A(T)$, where
$\alpha_A(T)$ is the total recombination coefficient
to all bound levels and 
$\alpha_B(T)$ is the recombination coefficient to all {\it excited} levels.
Hence, we have to integrate the radiation transfer equation 
including the emissivity by scatterings. In this frequency range,
the frequency-dependence of the opacity and the emissivity 
are thought to be small, because $\Delta\nu_T$ is
ten times smaller than $\nu_L$ for $T=10^4 {\rm K}$. 
So, we approximate the opacity and emissivity to be constant in this
frequency range.
After the transfer equation is solved, we calculate the
angle averaged intensity $J_{\nu_L}$ in terms of the numerical solution
$I_{\nu_L}$.  Using $J_{\nu_L}$, we obtain the photoionization rate 
coefficient for this frequency range,
$k_{\rm HI}^{sca}$;
\begin{eqnarray}
k_{\rm HI}^{sca}\simeq 
4\pi\frac{J_{\nu_L}\sigma_{\nu_L}}{h\nu_L}\Delta\nu_T 
\label{eq:ion1sca}
\end{eqnarray}
Finally, by adding (\ref{eq:ion1abs}) to (\ref{eq:ion1sca}), 
we obtain the total ionization rate $k_{\rm HI}$;
\begin{equation}
k_{\rm HI} = k_{\rm HI}^{abs} + k_{\rm HI}^{sca}.
\end{equation}

The UV heating rate $\Gamma_{\rm HI}$ is also obtained in a similar way.
The result is,
\begin{eqnarray}
\Gamma_{\rm HI} &=& \Gamma_{\rm HI}^{abs} + \Gamma_{\rm HI}^{sca},\\
\Gamma_{\rm HI}^{abs} &=& \int_{\nu_L+\Delta \nu_T }^\infty \int \frac{I_\nu}{h\nu} \sigma_{\nu} (h\nu-h\nu_L)d\omega d\nu, \nonumber  \\
          & \simeq& \frac{I_{\nu_L}^{\rm in}\sigma_{\nu_L}}{h}\int
\frac{1}{3\tau_{\nu_L}(\omega)^{1+\left(\alpha-1 \right)/3}}\gamma(1+\left(\alpha-1\right)/3,\tau_{\nu_L}\left(\omega\right)) h\nu_L d\omega \nonumber \\
\;&\;&-h\nu_Lk_{\rm HI}^{abs}, \\
\Gamma_{\rm HI}^{sca} &\simeq& 4\pi J_{\nu_L}\sigma_{\nu_L}\Delta\nu_T
\end{eqnarray}

In particular, for an infinite sheet, we have an asymptotic expression
for the absorption part ($\Gamma_{\rm HI}^{abs}$) 
in an optically thick limit as
\begin{equation}
\Gamma_{\rm HI}^{abs}\simeq 
\frac{2\pi I_{\nu_L}^{\rm in}\sigma_{\nu_L}\nu_L}{3}
\frac{\Gamma\left(1+\left(\alpha-1\right)/3\right)}{2+\left(\alpha-1\right)/3}
\tau_{\nu_L,v}^{-1-\left(\alpha-1\right)/3},
\label{eq:asymheat}
\end{equation} 
where $\Gamma(a)$ represents the gamma function.
It should be noted that the UV heating rate is not proportional to $\exp
(-\tau_{\nu_L,v})$, but is proportional to
$\tau_{\nu_L,v}^{-1-\left(\alpha-1\right)/3} $ for large $\tau_{\nu_L,v}$.
If we choose $\alpha=1$ as in the present paper, the UV heating rate is
proportional to $\tau_{\nu_L,v}^{-1}$. 
This shallow dependence upon the optical depth comes from a steep 
dependence of ionization cross section upon frequencies.
\newpage

\clearpage
\begin{figure}
\plotone{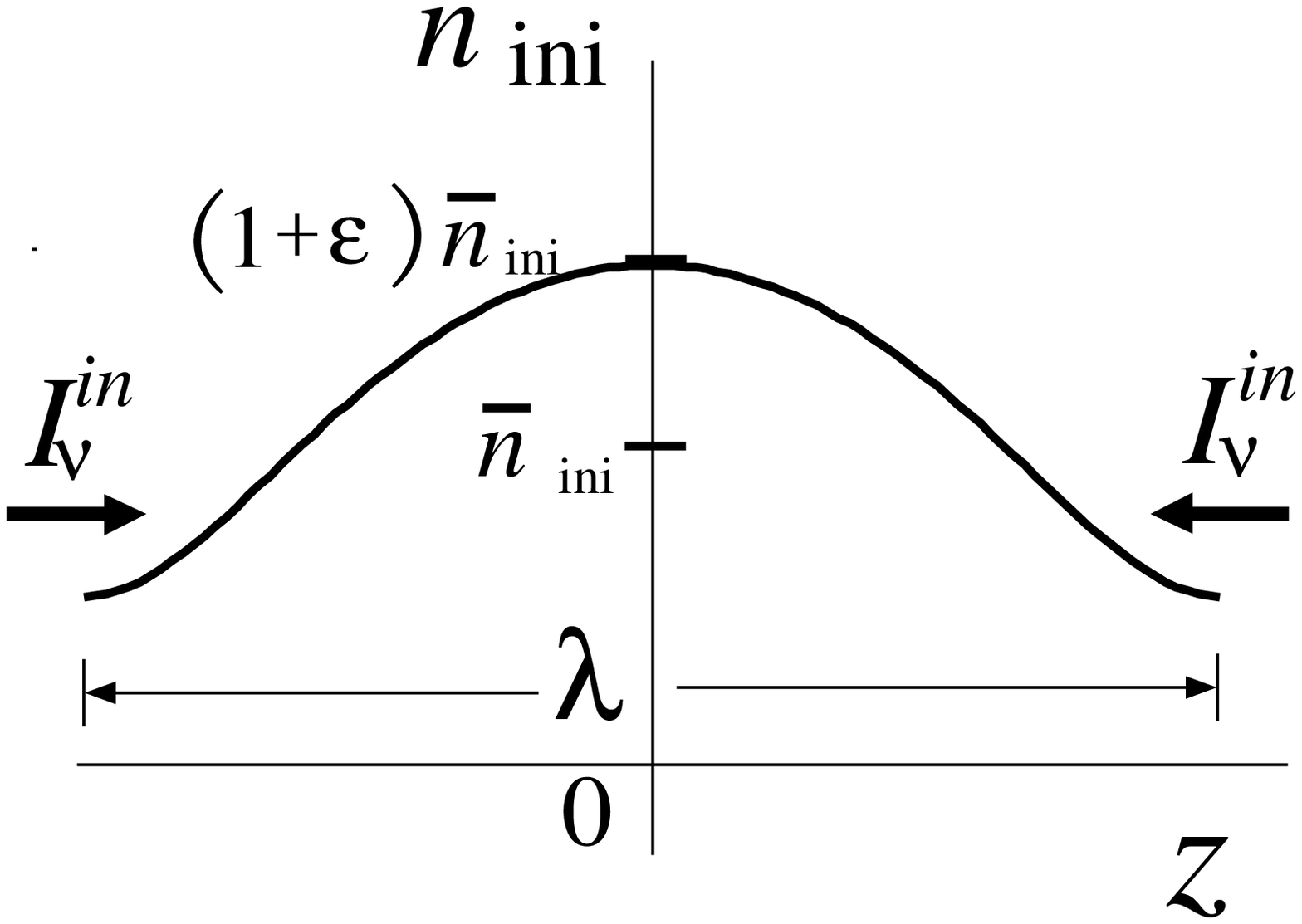}
\caption[f1.eps]{The profile for the initial density distributions. 
The horizontal axis, $z$, is the height form the midplane. The vertical
axis is the number density, $n_{\rm ini}$. 
$\bar{n}_{\rm ini}$ and $\lambda$ denote the mean number density
of the slab and the wave length of the slab, respectively. $\varepsilon$ is
the parameter which characterizes the initial density contrast
(see the text).}
\label{fig:cosin}
\end{figure}
\begin{figure}
\plotone{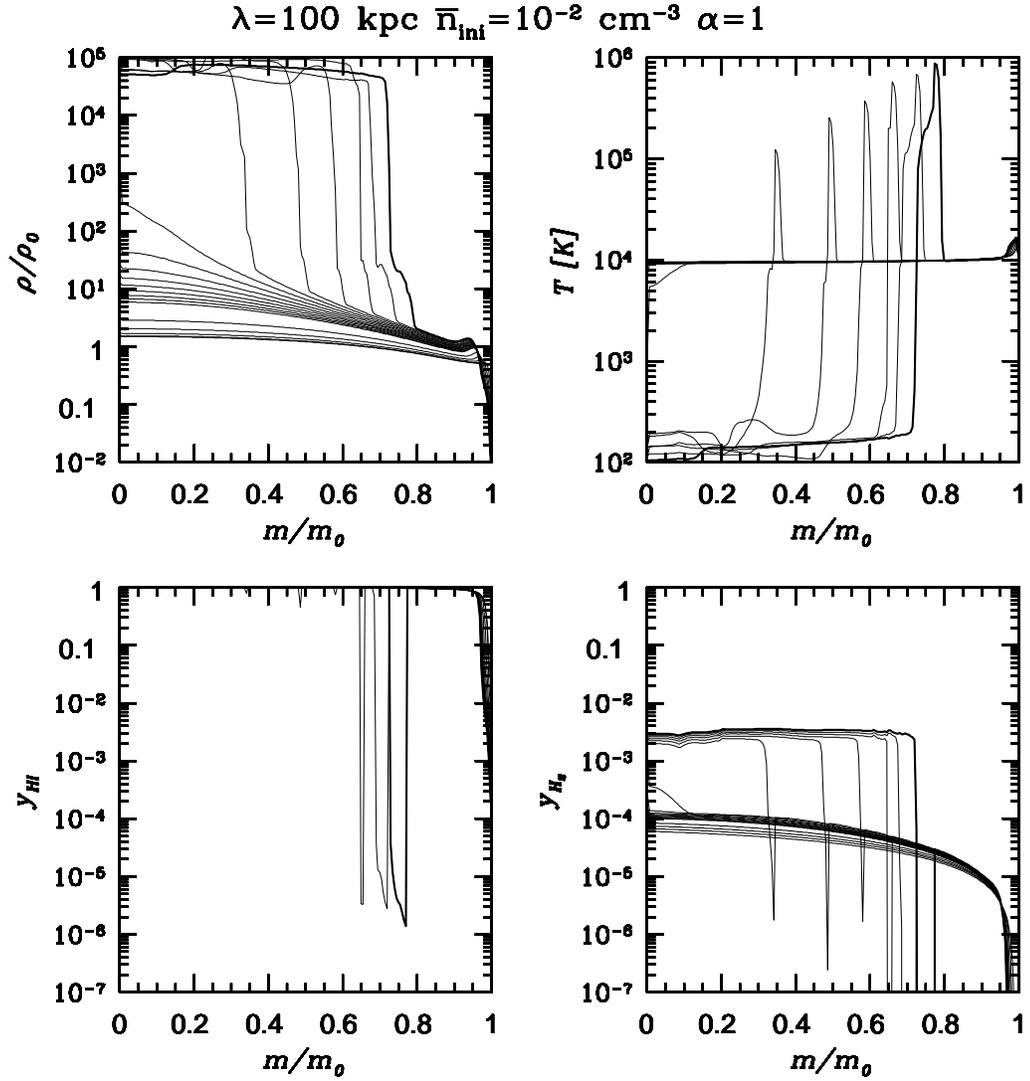}
\caption[f2.eps]{The time evolution of the spatial distributions 
for various physical quantities
is presented in the case of  $\bar{n}_{\rm ini}= 10^{-2}{\rm cm^{-3}} $,
 $\lambda=100$ kpc, and $\alpha=1$. 
The horizontal axis in each panel is the Lagrange
 coordinate normalized by the half total surface density, $m_0$.
Note that $m/m_0=0$
corresponds to the midplane of the sheet and $m/m_0=1$ is the surface.
The upper left panel shows the density distributions 
and the upper right panel shows the temperature distributions.
The lower left is the distributions for HI fraction, and
the lower right is the distributions for ${\rm H_2}$ fraction.
The thick solid lines show the final stage, i.e., 
$t=1.28 \times (4\pi G \bar{\rho}_{\rm ini})^{-1/2}$, 
where $\bar{\rho}_{\rm ini}=m_p \bar{n}_{\rm ini}$.
}
\label{fig:dist1}
\end{figure}
\begin{figure}
\plotone{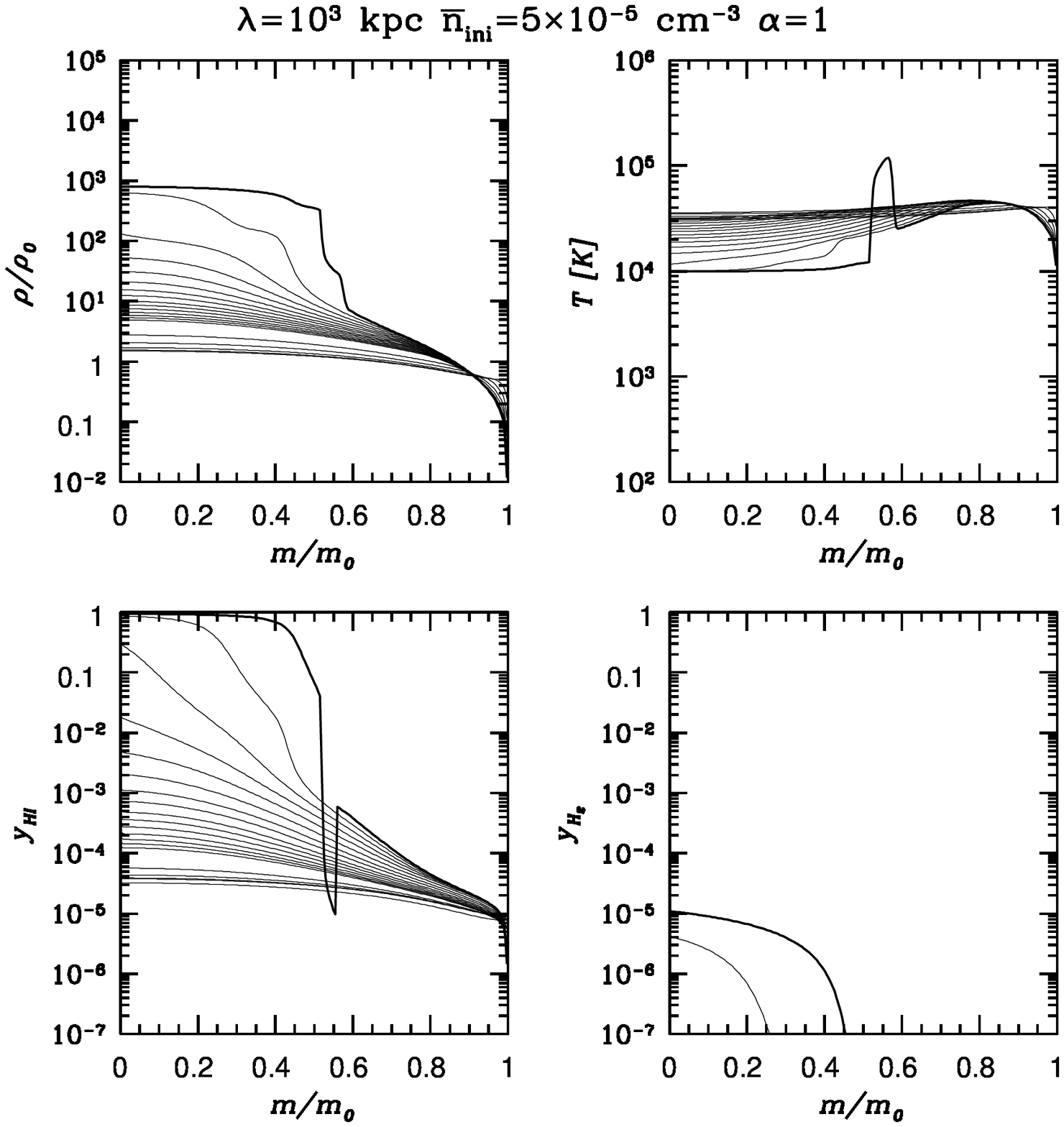}
\caption[f3.eps]{Same as Figure 2, except that 
$\bar{n}_{\rm ini}= 5\times10^{-5}{\rm cm^{-3}} $,
 $\lambda=1$ Mpc.}
\label{fig:dist2}
\end{figure}
\begin{figure}
\plotone{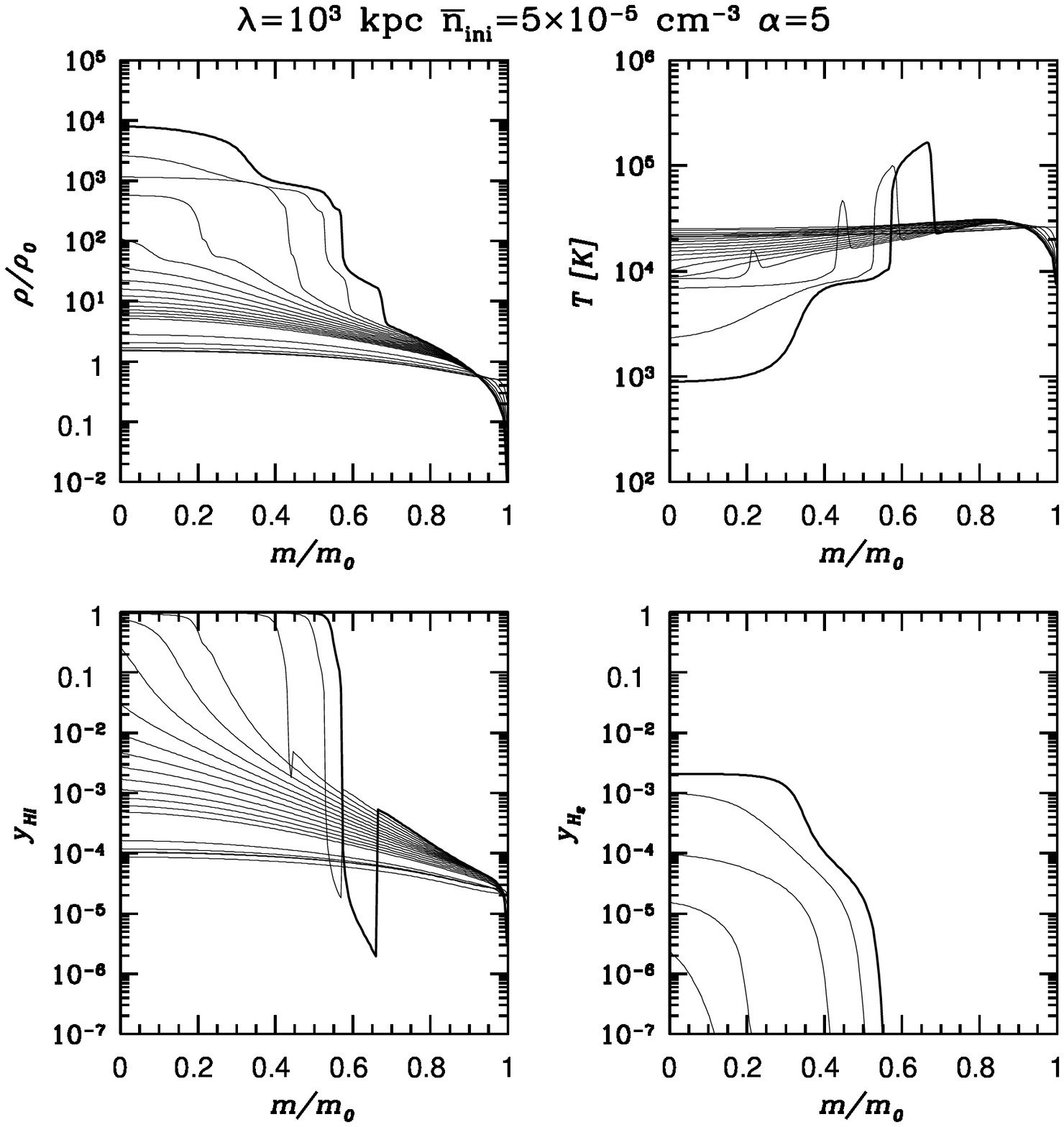}
\caption[f4.eps]{Same as Figure 2, except that 
$\alpha=5$.}
\label{fig:dist3}
\end{figure}
\begin{figure}
\plotone{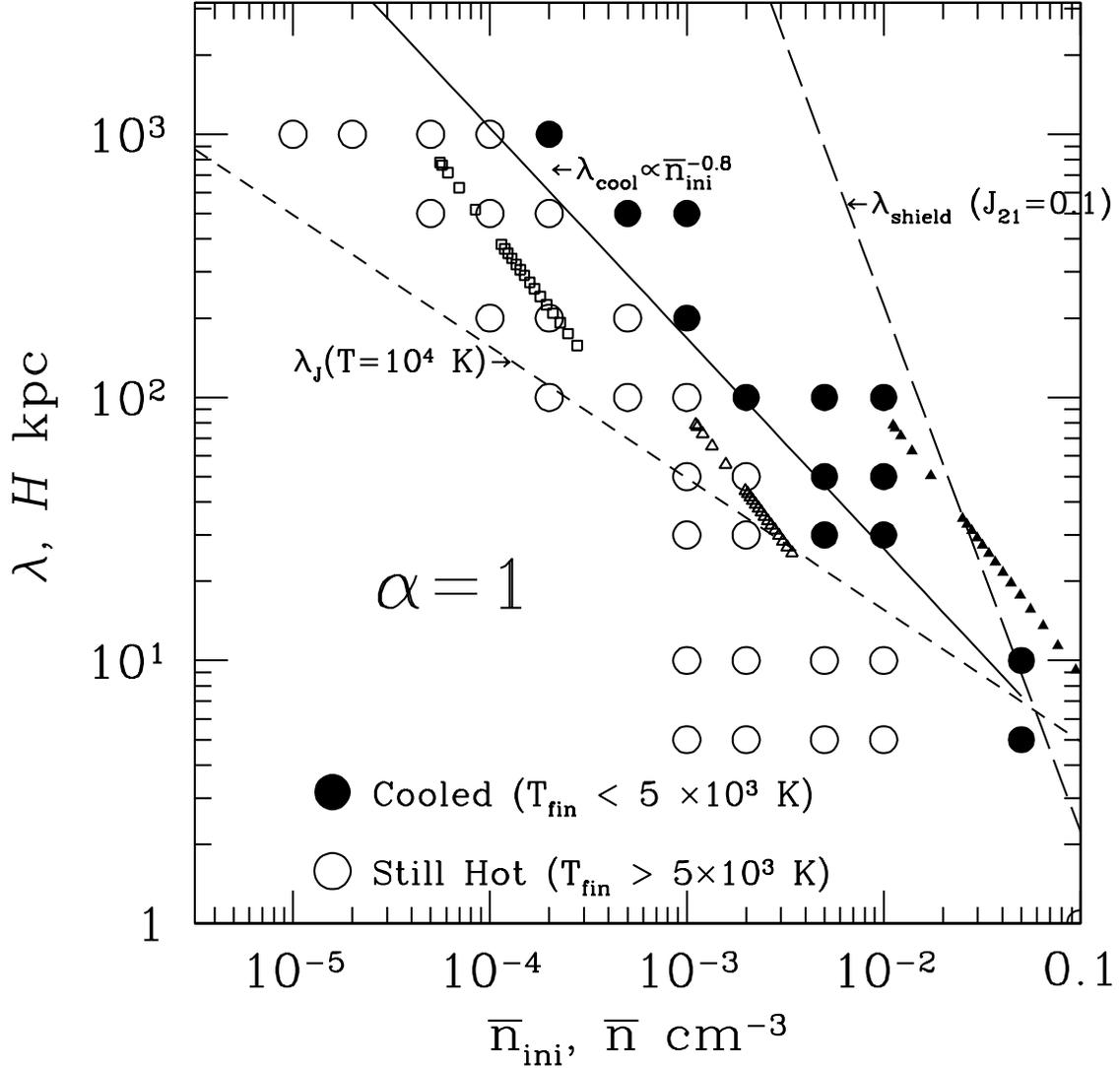}
\caption[f5.eps]{The physical states at $t=1.28/\sqrt{4\pi G\rho_{\rm ini}}$
 with given initial parameters are
 summarized. The vertical axis and the horizontal axis denote the
 thickness and the mean density of the sheets.  
The open and filled circles represent the initial conditions 
for the runs which we performed.
The open circles denote the sheets whose central temperature 
is high ($T > 5\times 10^3$ K) after one dynamical time.
The sheets denoted by the filled circles have
 cooled below $ 5\times 10^3$ K. Solid line ($\lambda_{\rm cool}$)
 denotes the boundary of the cold sheets and the hot ones. 
Long dashed line ($\lambda_{\rm shield}$) 
denotes the critical thickness above 
which the sheets are initially self-shielded against external UV
 heating. Short-dashed line is the Jeans length with $T=10^4$ K. 
Small open squares, open triangles, and filled triangles represent the
 evolutionary sequences for three initial parameters (see text). 
} 
\label{fig:collapse}
\end{figure}
\begin{figure}
\plotone{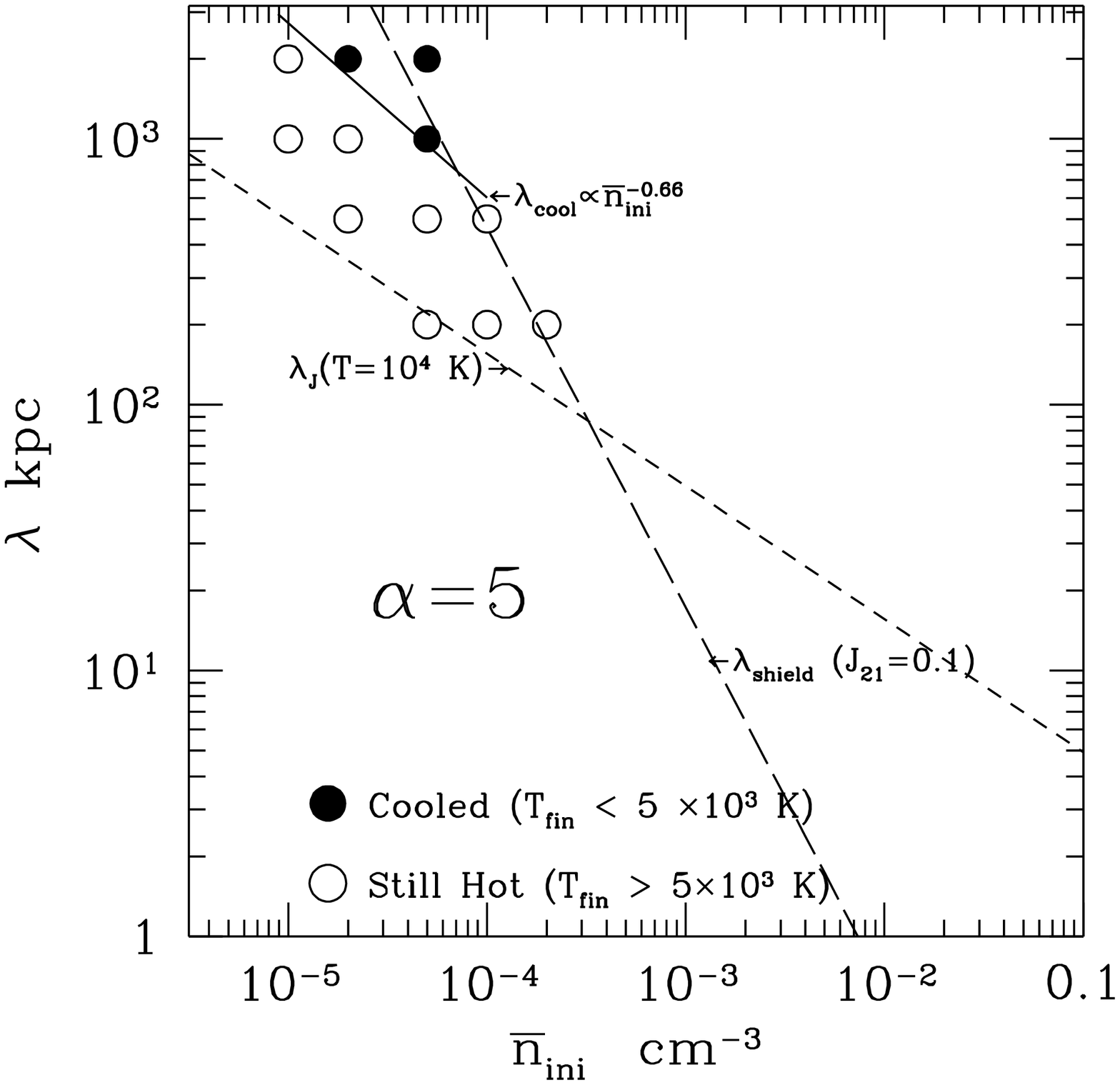}
\caption[f6.eps]{Same as Figure 5, except $\alpha=5$.
} 
\label{fig:collapse5}
\end{figure}

\begin{figure}
\plotone{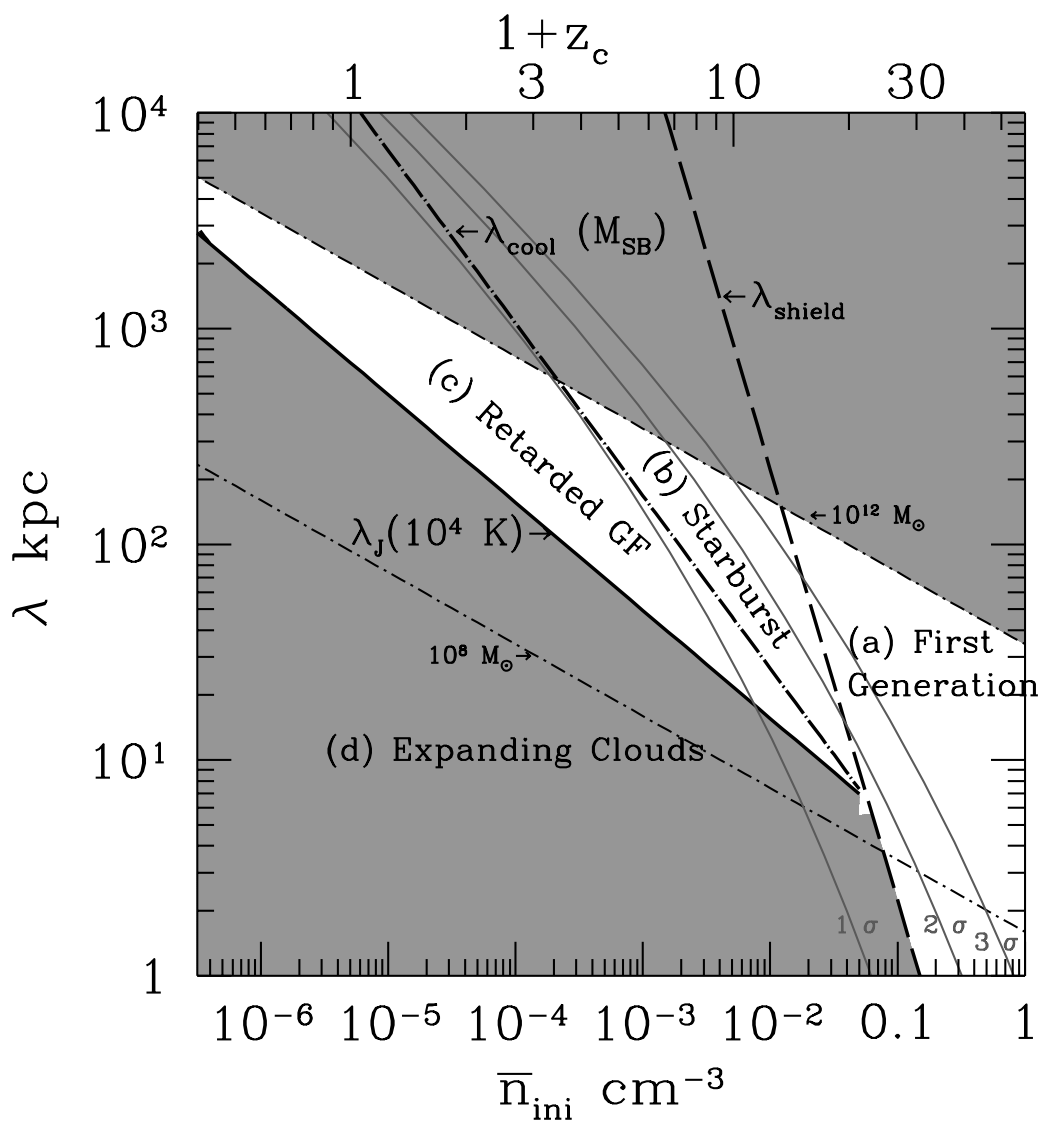}
\caption[f7.eps]{The initial parameter space is divided  
according to the evolution of clouds. Here,
$\alpha=1$ is assumed. Solid line, long-dashed line, dot-dashed line,
and short-dashed line
are the same as those in Figure 5. 
The upper abscissa is the estimated 
redshifts of the final collapse, $z_c$.
The dot-dashed line can be translated into 
$M_{\rm SB} = 2.2\times 10^{11} M_\odot \left[(1+z_c)/5\right]^{-4.2}$,
if the initial stage is assumed to be at the maximum expansion.
Thin dot-dashed lines 
denote the constant mass corresponding to $M=10^8$ and $10^{12} M_{\odot}$.
Assuming a standard CDM cosmology, 1$\sigma$, 2$\sigma$, and 3$\sigma$
density fluctuations are also shown.
The evolution of pregalactic clouds under UV background
radiation branches off into four categories; (a) first generation (promptly
self-shielded clouds), (b) starburst (initial starburst pancakes), (c) retarded
GF (retarded star formation in primordial galaxies), 
and (d) expanding clouds (unbound UV-heated clouds). 
}
\label{fig:galclass}
\end{figure}

\newpage
\def\RA{$\rightarrow$}
\def\Hbun{H$_2$}
\def\Hbunp{H$_2^+$}
\def\Hp{H$^+$}
\def\Hm{H$^-$}
\begin{deluxetable}{cc}
\tablecaption{Reaction rates \label{tab:reactions}}
\tablehead{\colhead{Reactions} & \colhead{References}}
\startdata
\Hp + e \RA H + $\gamma$ & Spitzer 1956 \markcite{Spitz} \nl
H + e \RA \Hm + $\gamma$ & de Jong 1972 \markcite{dejong} \nl
\Hm + H \RA \Hbun +e & Beiniek 1980 \markcite{Beiniek} \nl
3H \RA \Hbun + H& Palla, Salpeter \&  Stahler 1983 \markcite{PSS83}\nl
\Hbun + H \RA 3H & SK \markcite{SK87} \nl
2H + \Hbun \RA 2\Hbun& Palla, Salpeter \& Stahler 1983\markcite{PSS83}\nl
2\Hbun \RA 2H + \Hbun & SK \markcite{SK87}\nl
H + e \RA \Hp +2e & Lotz 1968 \markcite{Lotz}\nl
2H \RA H + \Hp + e & Palla, Salpeter \& Stahler 1983 \markcite{PSS83}\nl
H + \Hp \RA \Hbunp + $\gamma$ & Ramaker \& Peek 1976 \markcite{Ramaker}\nl
\Hbunp + H \RA \Hbun + \Hp & Karpas, Anicich \& Huntress 1979 \markcite{Karpas}\nl
\Hbun + \Hp \RA \Hbunp + H & Prasad \& Huntress 1980 \markcite{Prasad}\nl
\Hm + \Hp \RA \Hbunp + e &Poularert et al. 1978 \markcite{Poulaert}\nl
\Hbunp + e \RA 2H & Mitchell \& Deveau 1983 \markcite{Mitchell}\nl
\Hbunp + \Hm \RA H + \Hbun & Prasad \& Huntress 1980 \markcite{Prasad} \nl
\Hm + e \RA H + 2e &Duley 1984 \markcite{Duley}\nl
\Hm+ H \RA 2H + e &Izotov \& Kolensnik 1984 \markcite{Izotov}\nl
\Hm + \Hp \RA 2H &Duley 1984 \markcite{Duley}\nl
H + $\gamma$ \RA \Hp + e & see text \nl
\Hbun + $\gamma$ \RA 2H & Draine \& Bertoldi 1996 \markcite{DB96}\nl
\enddata
\end{deluxetable}
\begin{deluxetable}{cc|cccc}
\tablecaption{Initial physical parameters and the properties of cylindrical fragments \label{tab:frag} }
\tablehead{
\colhead{$\lambda\; {\rm kpc}$} & \colhead{$\bar{n}_{\rm ini}\;{\rm {cm}^{-3}}$}& \colhead{$l_0 \; M_\odot {\rm pc^{-1}}$} & \colhead{$f$} & \colhead{$x_s$ (shocked)} & \colhead{$x_c$ (cooled)}}
\startdata
100&$1\times 10^{-3}$ &$8.7\times 10^{3}$
&$2.4\times 10^{-1}$   & $0.54$ & not cooled \nl
100&$2\times 10^{-3}$&$6.3\times 10^{2}$&$1.7\times 10^{-1}$& $0.78$ &$0.47$ \nl
100&$1\times 10^{-2}$&$1.5\times 10^{1}$&$2.2\times 10^{-2}$& $0.36$ &$0.31$ \nl
\enddata
\end{deluxetable}

\end{document}